\documentclass[reprint]{revtex4-1}

\usepackage{hyperref}
\usepackage[utf8]{inputenc} 
\pdfoutput=1

\usepackage{amsmath,color,graphicx, amssymb, amsthm, mathtools, mathrsfs,caption,subcaption,natbib} 
\usepackage[section]{placeins}
\usepackage{enumerate}
\newtheorem{thm}{Theorem}

\theoremstyle{definition}

\theoremstyle{remark}

\theoremstyle{remark}

\usepackage{url}

\newcommand\mult{\operatorname{mult}}

\begin{document}
\title[Localization in \v{S}eba billiards]{Localization of low-energy eigenfunctions in \v{S}eba billiards}
\author{Minjae Lee}
\email{lee.minjae@math.berkeley.edu}
\address{Department of Mathematics, University of California, Berkeley}
\date{\today}

\begin{abstract}
We investigate localization of low-energy modes of the Laplacian with a point scatterer on a rectangular plate. We observe that the point scatterer acts as a barrier confining the low-level modes to one side of the plate while assuming the Dirichlet boundary condition at a point does not induce this type of localization. This low-energy phenomenon extends to higher modes as we increase the eccentricity of the plate.\end{abstract}
\maketitle

\section{Introduction}
Localization of modes in different physical systems is an interesting and puzzling phenomenon. It can be generated by the underlying geometry or by randomness. In this paper we consider the case in which localization is induced by the presence of a point scatterer on a two-dimensional plate, which is deterministic but requires either renormalization or spectral theory to be properly defined.

The specific model is called the \v{S}eba billiard and was introduced in \cite{Seba} to study quantum chaos. See also \cite{Shigehara1,Shigehara2,rudnick, Ueberschar} for further developments. \v{S}eba considered a limiting case of a standard model of ergodic dynamics, the Sinai billiard \cite{Sinai}, which is a rectangle with a disk removed.  In the standard quantization of that model one considers the Laplace operator with zero (Dirichlet) boundary conditions on the boundaries of the rectangle and of the disk. In the \v{S}eba model the disk is shrunk to a point with a suitable renormalization. That renormalization can be interpreted as a choice of a self-adjoint extension \cite{reedsimon2} of the Laplacian on the rectangle with the point removed. The point is then called a {\em point scatterer}. We show that the presence of such a scatterer has a dramatic effect on the localization of low-lying modes. For other two-dimensional structures, a localization for the modes of the Laplacian was studied Sapoval \emph{et al.} \cite{fractal, irregular, review1} in the case of irregular geometry or fractal boundaries. Filoche and Mayboroda \cite{bilaplacian} discovered that localization can be achieved for modes of the bi-Laplacian $ \Delta^2$ on a rectangle with a point removed. For this fourth order operator the natural boundary conditions require the mode and its gradient to vanish at the boundary. Physically, this boundary condition means that the plate is {\em clamped} at the boundary and at the interior point.

In \cite{bilaplacian}, numerical analysis of the modes of the bi-Laplacian showed strong localization on one side of the clamped interior point. Somewhat surprisingly, the same phenomenon occurs for the \v{S}eba billiard, that is, for a model with quantum mechanical origins. As pointed out in \cite{bilaplacian} this phenomenon does occur for limits of eigenfunctions on Sinai billiards with shrinking disks. In our
language that means that localization does not occur without renormalization.

\section{Formalism}
Point scatterers are formally defined by a Schr\"{o}dinger operator \(-\Delta+c\delta_{\mathbf{x}_0}\) where \(c\) is constant and \(\delta_{\mathbf{x}_0}\) is the Dirac delta function located at a specific point \(\mathbf{x}_0\). More precisely, it is a self-adjoint extension of the Laplacian whose domain consists of the functions vanishing at \(\mathbf{x}_0\). A point scatterer in a rectangle with the Dirichlet boundary condition is called the \v{S}eba billiard \cite{Seba}. 

Consider a rectangle \(\Omega = [0,a] \times [0,b]\) with \(a, b>0\) and the Dirichlet Laplacian \[-\Delta : H^2(\Omega)\cap H_0^1(\Omega) \rightarrow L^2(\Omega).\] Then we have the eigenvalues \(0<E_1\le E_2\le\cdots\) of \(-\Delta\) with the corresponding \(L^2\)-normalized eigenfunctions \(\phi_1,\phi_2, \cdots\).

On the other hand, we construct a point scatterer at \(\mathbf{x}_0 =(x_0,y_0)\in \Omega\) as follows: First, restrict the domain of the Dirichlet Laplacian \(-\Delta\) to the functions vanishing at \(\mathbf{x}_0\in \Omega\). By the theory of self-adjoint extension developed by von Neumann, such a symmetric operator has a family of self-adjoint extensions \(-\Delta_{\alpha,\mathbf{x}_0}\) with a parameter \(\alpha \in (-\infty,\infty ]\). More precisely, let \(G_z\) be the integral kernel of the resolvent \((-\Delta-z)^{-1}: L^2(\Omega) \rightarrow L^2(\Omega)\), namely,
\[G_z(\mathbf{x},\mathbf{x}') = \sum_{n=1}^\infty \frac{\phi_n(\mathbf{x}) \phi_n(\mathbf{x}')}{E_n-z}\]
so that for \(f \in L^2 (\Omega)\),
\[(-\Delta-z)^{-1} f (\mathbf{x}) = \int_\Omega G_z(\mathbf{x},\mathbf{x}') f(\mathbf{x}') d\mathbf{x}'.\]

Then for \(z\in \rho (-\Delta_{\alpha,\mathbf{x}_0})\), the integral kernel of \((-\Delta_{\alpha,\mathbf{x}_0}-z)^{-1} : L^2(\Omega)\rightarrow L^2(\Omega)\) reads
\begin{multline}\label{dom}
(-\Delta_{\alpha,\mathbf{x}_0}-z)^{-1}(\mathbf{x},\mathbf{x}') \\= G_z(\mathbf{x},\mathbf{x}') + \left[\alpha-F(z) \right]^{-1} G_z(\mathbf{x}_0,\mathbf{x}') G_z(\mathbf{x},\mathbf{x}_0)
\end{multline}
where
\begin{equation}\label{Fz}
F(z)=\sum_{n=1}^\infty \phi_n(\mathbf{x}_0)^2\left( \frac{1}{E_n-z}-\frac{E_n}{E_n^2+1} \right) \end{equation}
(see Fig.~\ref{Fz}).


The coupling constant \(\alpha\in (-\infty,\infty]\) can be considered a parameter related to the strength of the point scatterer. Note that the point scatterer annihilates as \(\alpha \rightarrow\pm \infty\) whereas it acts stronger when \(|\alpha|\ll\infty\).

Now we consider the spectral property of \v{S}eba billiards. Let \(\sigma(P)\) denote the spectrum of an operator \(P\) and let \(\mult(z,P)\) denote the multiplicity of an eigenvalue \(z\in \sigma(P)\).
As the Dirichlet Laplacian \(-\Delta\) has a purely discrete spectrum, so does \(-\Delta_{\alpha,\mathbf{x}_0}\). In addition, some eigenvalues of \(-\Delta_{\alpha,\mathbf{x}_0}\) remain in \(\sigma(-\Delta)\) regardless of the coupling constant \(\alpha\) while the others do not. Hence, for \(\alpha\in\mathbb{R},\) we divide \(\sigma(-\Delta_{\alpha,\mathbf{x}_0})\) into the following two types: 
\begin{enumerate}
\item Perturbed eigenvalues: \(\sigma(-\Delta_{\alpha,\mathbf{x}_0})\setminus\sigma(-\Delta) \) and
\item Unperturbed eigenvalues: \(\sigma(-\Delta_{\alpha,\mathbf{x}_0})\cap\sigma(-\Delta)\)
\end{enumerate}
where each of them is obtained by different conditions as follows:
\begin{thm}\label{pert}
For \(\alpha\in \mathbb{R}\), \(z\in\sigma(-\Delta_{\alpha,\mathbf{x}_0})\setminus\sigma(-\Delta)\) if and only if \[\alpha=F(z).\]
Then \(\mult(z,-\Delta_{\alpha,\mathbf{x}_0})=1\) with the corresponding eigenfunctions \[\psi(\mathbf{x})=N^{-1} G_z(\mathbf{x},\mathbf{x}_0)\] where \(N=\|G_z(\bullet,\mathbf{x}_0)\|_{L^2(\Omega)}\) is the normalization constant.
\end{thm}
\begin{figure}
\centering
\includegraphics[width=60mm]{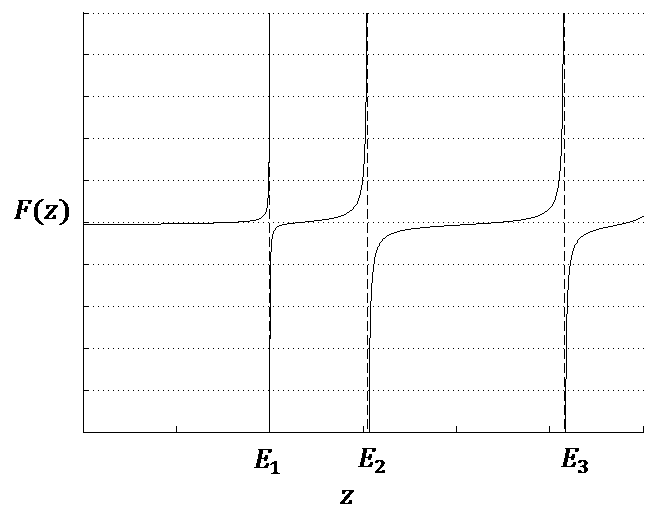}
\caption{A schematic graph of \(F(z)\) defined by Eq.\eqref{Fz}.}
\label{Fz}
\end{figure}

\begin{thm}\label{unpert}
Define \(\mu\) and \(\mu_0\) as
\begin{align}\mu(z)&\equiv\mult(z,-\Delta)=\#\{n\ge1 ~|~ z=E_n\}
\\ \mu_0(z) &\equiv \#\{n\ge1 ~|~ z=E_n, \phi_n(\mathbf{x}_0)=0\}.\end{align}

Then for \(\alpha\in \mathbb{R},~z\in \sigma(-\Delta_{\alpha,\mathbf{x}_0})\cap\sigma(-\Delta) \) if and only if 
\[\mu_0 (z) \ge 1 ~\text{ or }~ \mu(z)\ge 2\]
Also,
\[\mult(z,-\Delta_{\alpha,\mathbf{x}_0})=\begin{cases}
\mu(z), & \mbox{ if }\mu_0(z) =\mu(z)\\
\mu(z)-1, & \mbox{ if } \mu_0(z) <\mu(z)
\end{cases}\] with the corresponding eigenspaces
\[\left\{\sum_{z=E_n} c_n\phi_n ~\middle|~ \sum_{z=E_n} c_n\phi_n(\mathbf{x}_0)=0, \quad c_n\in\mathbb{C}\right\}.\]
\end{thm}

Proofs can be found in Chapter 2 of \cite{pseudolaplacian} with generalized statements for a compact Riemannian manifold of dimension two or three. The coupling constant \(\alpha\)  in Eq.\eqref{dom} can be obtained by following the notations provided by Albeverio \emph{et al.} \cite{solvable}. Note that \(\alpha\) also corresponds to the inverse of the coupling constant \(v_B\) or \(\overline{v}_\theta\) in Shigehara's setting \cite{Shigehara1,Shigehara2}.

We may interpret Theorem~\ref{unpert} as that the Laplacian eigenfunctions vanishing at \(\mathbf{x}_0\) do not feel the presence of the point scatterer. So not only do they remain as the eigenfunctions of \(-\Delta_{\alpha,\mathbf{x}_0}\), but also the associated eigenvalues stay in \(\sigma(-\Delta_{\alpha,\mathbf{x}_0})\) for any \(\alpha\).

On the other hand, by combining Theorem~\ref{pert} and \ref{unpert} we obtain that the eigenvalues of the point scatterer are interlaced between those of the Dirichlet Laplacian. In other words, for \(\alpha\in (-\infty,\infty]\), let \(z_1(\alpha)\le z_2(\alpha)\le \cdots \) be the eigenvalues of \(-\Delta_{\alpha,\mathbf{x}_0}\). Then we have
\[z_1(\alpha) \le E_1 \le z_2(\alpha) \le E_2 \le z_3(\alpha) \le E_3 \le \cdots.\]
In addition, for \(n \ge 1\),
\begin{align*}\lim_{\alpha\rightarrow\infty}z_n(\alpha) &= E_n \\
\lim_{\alpha\rightarrow -\infty}z_{n+1}(\alpha) &= E_n\\
\lim_{\alpha\rightarrow -\infty}z_{1}(\alpha) &=-\infty\end{align*}

\section{Localization of Eigenfunctions}\label{result}
In this section, we show several examples of perturbed eigenfunctions localized on a plate due to the point scatterer with a suitable coupling constant \(\alpha\in\mathbb{R}\).

\begin{figure}
\centering
\includegraphics[width=60mm]{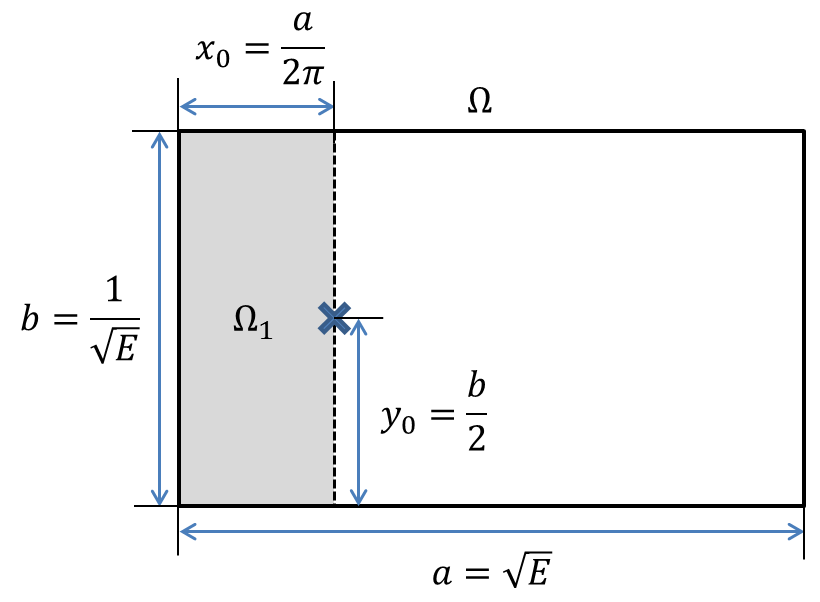}
\caption{(Color online) Geometry of a point scatterer at \(\mathbf{x}_0=(x_0,y_0) \) (marked as \(\times\)) in \(\Omega\). The left part of the plate divided by \(\mathbf{x}_0\) is denoted by \(\Omega_1=[0,x_0]\times[0,b]\). }
\label{plate}
\end{figure}

Let \(\Omega=[0,a]\times [0,b]\) with \(a=\sqrt{E}\) and \(b=1/\sqrt{E}\) so every plate has unit area for any \(E>0\) which is the eccentricity of the plate. The unperturbed eigenfunctions obtained by Theorem~\ref{unpert} are independent of \(\alpha\) so they have no chance to be localized at all. In order to avoid such cases as much as possible, first we assume the eccentricity \(E\) to be irrational so that all \(E_n\)'s are nondegenerate. In addition, let \(\frac{a}{x_0}\) be irrational to minimize the case in which \(\phi_n\) vanishes at \(\mathbf{x}_0\).
In this paper, we choose a specific value \(\frac{a}{x_0}=2\pi\) (Fig.~\ref{plate}). However, it should be noted that the qualitative property we observe also holds for other values of \(\frac{a}{x_0}\) as long as they are irrational.

By Theorem~\ref{pert}, if \(z_n(\alpha)\) is a perturbed eigenvalue of \(-\Delta_{\alpha,\mathbf{x}_0}\) then the corresponding normalized eigenfunction \(\psi_{n,\alpha} \in L^2(\Omega)\) satisfies the following \(L^2\)-identity:
\begin{equation}\label{psina}
\psi_{n,\alpha}(\mathbf{x}) = N_{n,\alpha}^{-1}\sum_{n'=1}^\infty \frac{\phi_{n'}(\mathbf{x}) \phi_{n'}(\mathbf{x}_0)}{E_{n'}-z_n(\alpha)}
\end{equation}
where \(N_{n,\alpha}\) is the \(L^2\)-normalization constant.

We now investigate the localization of the perturbed eigenfunctions given by Eq.\eqref{psina} which depends on the mode number \(n\), the coupling constant \(\alpha\), and the eccentricity \(E\). Among those three variables, we mainly concentrate on \(n\) and \(E\). It should be noted that \(\alpha\) is chosen to maximize the localization property for each situation. 

In order to quantify the localization of multiple modes with ease, we introduce two kinds of measurement: First, we define the \(L^2\)-norm ratio \(R_1(n,\alpha)\) as
\begin{equation}\label{R1na}
R_1(n,\alpha) = \left(\int_{\Omega_1} |\psi_{n,\alpha}(\mathbf{x})|^2 d\mathbf{x}\right)^\frac{1}{2},
\end{equation}
where \(\Omega_1=[0,x_0]\times[0,b]\) denotes the left part of the plate divided by the point scatterer.
In addition, let \(A(n,\alpha)\) be the amplitude at \(\mathbf{x}_0\): 
\begin{equation}\label{Ana}
A(n,\alpha)=|\psi_{n,\alpha}(\mathbf{x}_0)|.
\end{equation}
For simplicity, let us omit \(E\) in those notations since it is already embedded in every \(E_n\) and \(\phi_n\) of Eq.\eqref{psina}.

Note that we assume that all eigenfunctions are \(L^2\)-normalized. Then \(R_1(n,\alpha)\) measures the ratio of the \(L^2\)-norm localized in \(\Omega_1\). For instance, \(R_1(n,\alpha)=0\) and \(R_1(n,\alpha)=1\) imply that \(\psi_{n,\alpha}\) is completely localized in \(\Omega\setminus\Omega_1\) and \(\Omega_1\), respectively. On the other hand, \(A(n,\alpha)\) measures how much the point scatterer at \(\mathbf{x}_0\) attracts the amplitude of modes.

\begin{figure}
\centering
\begin{subfigure}{0.49\columnwidth}
\centering
\includegraphics[width=\textwidth]{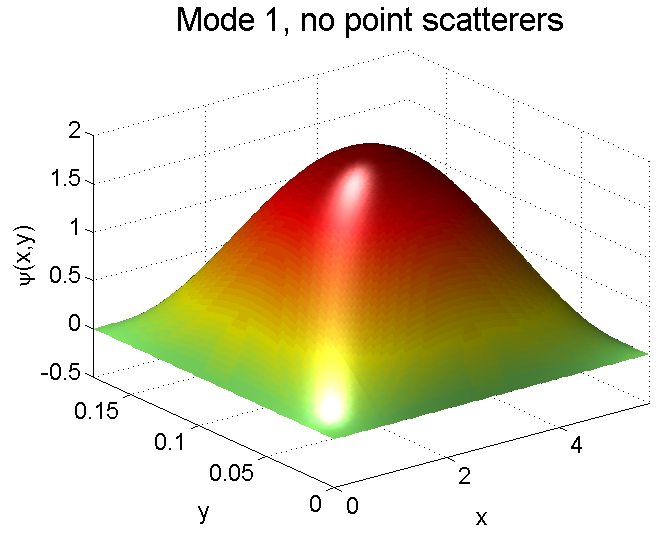} 
\end{subfigure}
\hfill
\begin{subfigure}{0.49\columnwidth}
\centering
\includegraphics[width=\textwidth]{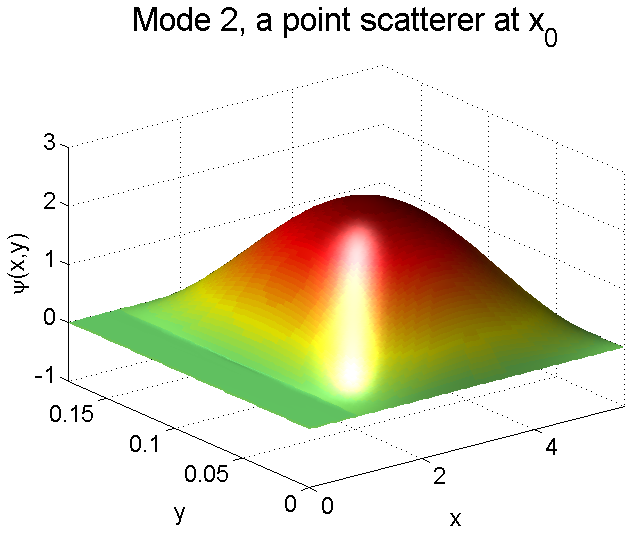} 
\end{subfigure}

\begin{subfigure}{0.49\columnwidth}
\centering
\includegraphics[width=\textwidth]{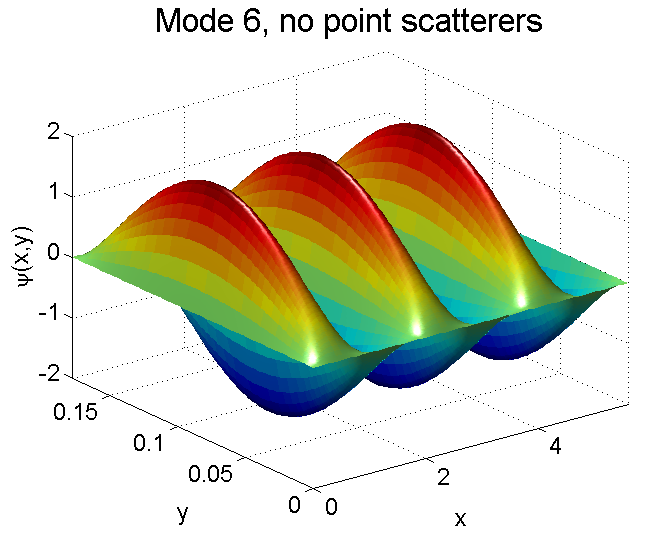} 
\end{subfigure}
\hfill
\begin{subfigure}{0.49\columnwidth}
\centering
\includegraphics[width=\textwidth]{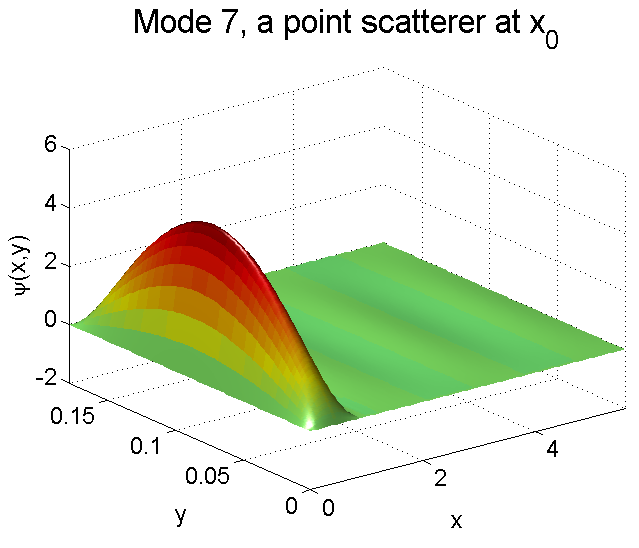} 
\end{subfigure}

\begin{subfigure}{0.49\columnwidth}
\centering
\includegraphics[width=\textwidth]{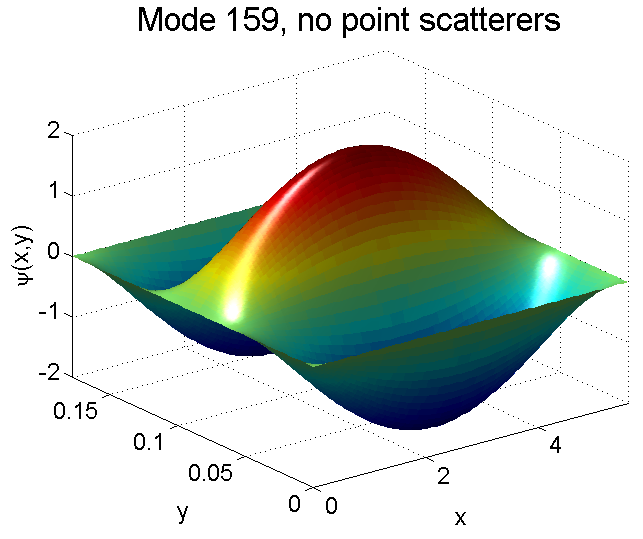} 
\end{subfigure}
\hfill
\begin{subfigure}{0.49\columnwidth}
\centering
\includegraphics[width=\textwidth]{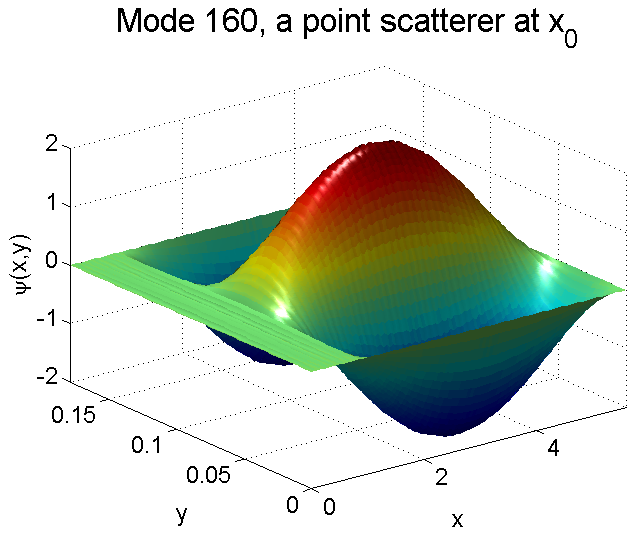} 
\end{subfigure}

\caption{(Color online) Several modes on a plate with eccentricity \(E=10\pi\). The figures on the left- and right-hand columns correspond to the Dirichlet Laplacian and a point scatterer at \(\mathbf{x}_0\), respectively. For the point scatterer, the coupling constant \(\alpha\) is chosen to maximize the localization. One can observe several modes localized on the left- or the right-hand side of the point scatterer.}\label{eigftn}
\end{figure}

\subsection{Point scatterer acting as a barrier}
Now we provide numerical results showing that the low-level eigenfunctions with \(n\ge 2\) localize to the left or the right of \(\mathbf{x}_0\) where the point scatterer is located. In Fig.~\ref{eigftn}, we compare some eigenfunctions localized by a point scatterer (right-hand column) to those of the Dirichlet Laplacian (left-hand column) where \(E=10\pi\). These modes are examples in which the point scatterer acts as a barrier confining the amplitude distribution to the left or right of itself.

Instead of presenting the amplitude distribution of every localized eigenfunction on the plate \(\Omega\), let us draw a graph of the \(L^2\)-norm ratio \(R_1(n,\alpha)\) as a function of the mode number \(n\) for each \(E\) fixed. The eigenfunction \(\psi_{n,\alpha}\) is considered to be localized in terms of the \(L^2\)-norm ratio if \(R_1(n,\alpha)<0.1\) or \(R_1(n,\alpha)>0.9\).

\begin{figure}
\centering
\begin{subfigure}{0.49\columnwidth}
\centering
\includegraphics[width=\textwidth]{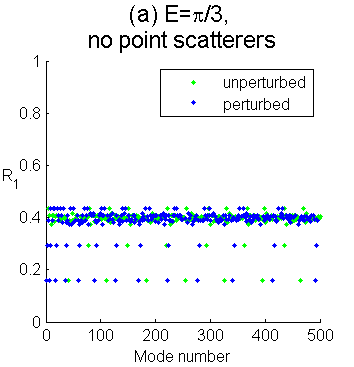} 
\end{subfigure} \hfill
\begin{subfigure}{0.49\columnwidth}
\centering
\includegraphics[width=\textwidth]{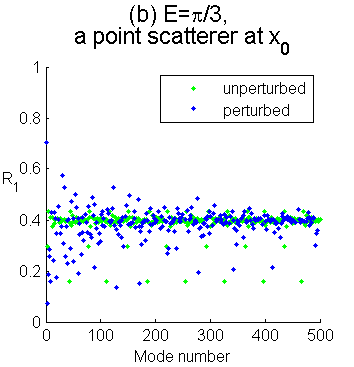} 
\end{subfigure}

\begin{subfigure}{0.49\columnwidth}
\centering
\includegraphics[width=\textwidth]{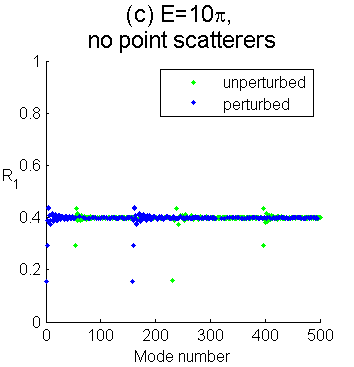} 
\end{subfigure} \hfill
\begin{subfigure}{0.49\columnwidth}
\centering
\includegraphics[width=\textwidth]{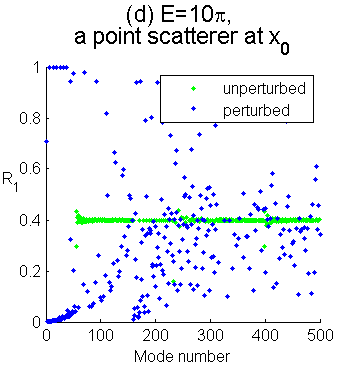} 
\end{subfigure}

\caption{(Color online) \(L^2\)-norm ratio \(R_1\) of the first 500 modes with a point scatterer at \(\mathbf{x}_0\). \(\alpha\) is chosen to maximize the number of localized modes.
One can observe that the \(L^2\)-norm ratio tends to polarize to either 0 or 1 for a strong (e.g. \(|\alpha|\ll\infty\)) point scatterer. Such a tendency appears remarkably to perturbed lower level modes in (d) as the eccentricity \(E\) increases.}\label{PR}
\end{figure}

Fig.~\ref{PR} compares \(R_1(n,\alpha)\) of the first 500 eigenfunctions of \(-\Delta_{\alpha,\mathbf{x}_0}\) to those of the Dirichlet Laplacian where \(E=\frac{\pi}{3}\) and \(E= 10\pi\). For each \(E\), \(\alpha\) is chosen to maximize the number of localized modes. The blue (black) points and green (gray) points represent the eigenvalues given by Theorem~\ref{pert} and \ref{unpert}, respectively. Note that if the modes are localized completely to the right or the left of \(\mathbf{x}_0\) then all points in the graph will be polarized to either 0 or 1. When eccentricity is small (\(E=\frac{\pi}{3}\)), the point scatterer weakly perturbs the \(L^2\)-norm ratio of modes but it is hard to say these modes are localized enough. On the other hand, when eccentricity is large (\(E=10\pi\)), one can observe a strong localization especially at the low-level modes. Video clips for the continuous transition of Fig.~\ref{PR} from the Dirichlet Laplacian to a point scatterer for \(E=\frac{\pi}{3}, ~E=10\pi\) are given in \url{http://math.berkeley.edu/~lmj0425/seba_PR_pi3.avi} and \url{http://math.berkeley.edu/~lmj0425/seba_PR_10pi.avi}, respectively. Note that the Dirichlet Laplacian is equivalent to the point scatterer with \(\alpha =\infty\).

Now we discuss how far the localization in terms of the \(L^2\)-norm ratio maintains its influence up to the higher-level eigenfunctions. It has been proved by Keating \emph{et al.}. \cite{local} that the eigenfunctions of \v{S}eba billiards tend to localize around eight points in momentum space as the level of the mode increases. In other words, the localization in position space we observe in this paper is an intermediate phenomenon that tends to diminish as the mode number increases. 

\begin{figure}
\centering
\includegraphics[width=0.8\columnwidth]{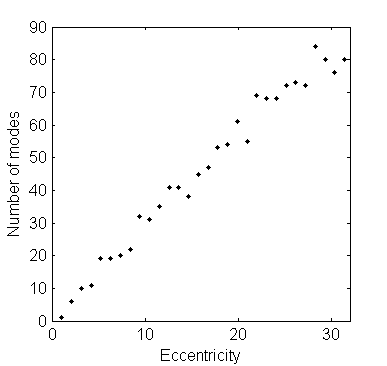} 
\caption{The number of localized modes (\(R_1 <0.1\) or \(R_1 >0.9\) ) out of the first 500 modes on the plate of eccentricity \(E\). For each \(E\), the coupling constant \(\alpha\) is chosen to maximize the number of localized modes.}
\label{locmodes_ecc}
\end{figure}

However, one can observe that the localization effect extends to higher-level eigenfunctions as the eccentricity \(E\) increases. Fig.~\ref{locmodes_ecc} displays the number of localized modes out of the first 500 modes as a function of eccentricity \(E\). The coupling constant \(\alpha\) is chosen to maximize the number of localized modes for each \(E\). Therefore, we can conclude that the point scatterer induces a strong localization as a barrier confining the amplitude of low-level eigenfunctions to either \(\Omega_1\) or \(\Omega\setminus\Omega_1\). 

\subsection{Point scatterer acting as an attractor}
On the other hand, the eigenfunction corresponding to the lowest eigenvalue \(z_1(\alpha) \in (-\infty,E_1)\) shows a different behavior: It tends to localize around \(\mathbf{x}_0\) so we can say the point scatterer attracts the amplitude of the first mode.

\begin{figure}
\centering
\includegraphics[width=0.8\columnwidth]{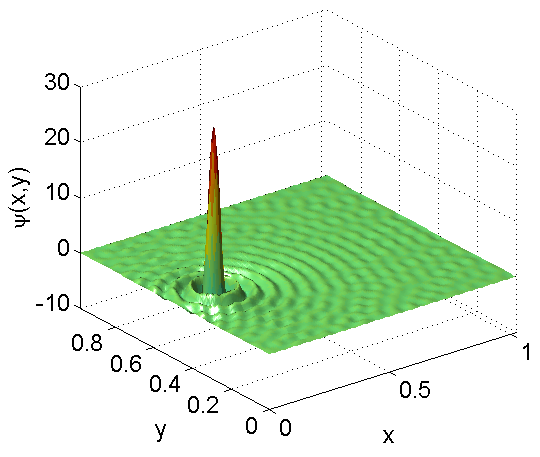}
\caption{(Color online) Mode 1 of a point scatterer at \(\mathbf{x}_0\) on a plate with \(E=\frac{\pi}{3}\) and \(\alpha=-0.48\). The associated eigenvalue is \(z_1(\alpha)=-1.29\times 10^4\). The amplitude is highly localized around the point scatterer but not biased to either the left or the right of it.}
\label{mode1}
\end{figure}

A numerical simulation indicates that the amplitude at \(\mathbf{x}_0\) mainly depends on the mode number. In particular, the first mode with the associated eigenvalue \(z_1(\alpha) \in (-\infty, E_1)\) tends to localize around \(\mathbf{x}_0\) as \(z_1(\alpha)\rightarrow -\infty\), or equivalently, as \(\alpha\rightarrow -\infty\). Fig.~\ref{mode1} shows the eigenfunction of \(-\Delta_{\alpha,\mathbf{x}_0}\) corresponding to \(z_1(\alpha)=-1.29\times 10^4\) with \(\alpha=-0.48\). Since the amplitude localizes around the point scatterer evenly, our first criterion using the \(L^2\)-norm ratio cannot detect this type of localization. So we introduce the second measurement \(A(n,\alpha)\), the amplitude of the mode at \(\mathbf{x}_0\), to investigate the behavior described above.

\begin{figure}
\centering
\includegraphics[width=0.95\columnwidth]{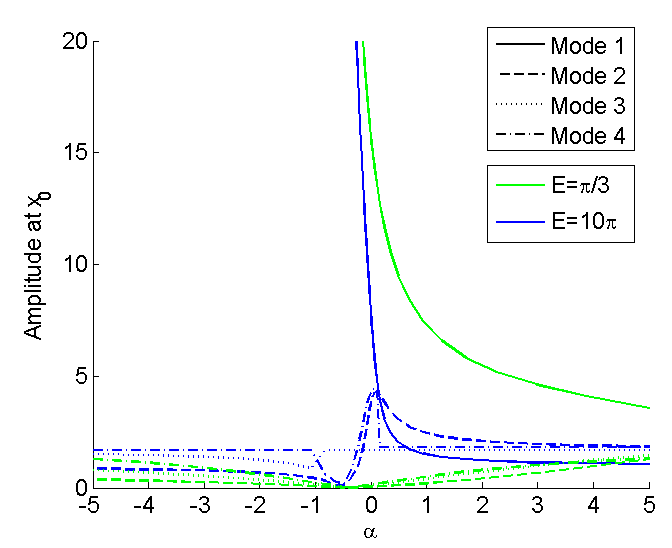}
\caption{(Color online) Amplitude of the first four modes of at \(\mathbf{x}_0\) for the coupling constant \(\alpha \in [-5,5]\). Green (gray) lines and blue (black) lines correspond to the eccentricities \(E=\frac{\pi}{3}\) and \(E=10\pi\), respectively. The first mode tends to localize around the point scatterer as \(\alpha\rightarrow -\infty\) regardless of the eccentricity while the others maintain low amplitude for all \(\alpha\)'s.}
\label{Amp_a}
\end{figure}

Fig.~\ref{Amp_a} displays how the presence of the point scatterer with the coupling constant \(\alpha\) affects \(A(n,\alpha)\) of the first four modes where \(E=\frac{\pi}{3}\) [green (gray) lines] and \(E=10\pi\) [blue (black) lines]. Regardless of the eccentricity, the amplitude of the first mode at \(\mathbf{x}_0\) blows up as \(\alpha\rightarrow -\infty\) but such localization does not occur in the other modes. This can be justified by the Fourier series representation of the perturbed eigenfunction in Eq.~\eqref{psina} since, for each \(\phi_{n}\), the Fourier coefficients \[\frac{\phi_{n}(\mathbf{x}_0)}{E_{n}-z}\] get relatively uniform as \(z\rightarrow -\infty\). On the other hand, if \(E_j\le z \le E_{j+1}\) for some \(j\ge 1\), then the Fourier coefficients corresponding to \(E_n\)'s near \(z\) prevail in the summation which prevents the amplitude of higher modes from diverging at a certain point.

\section{Conclusion}
We have shown that the point scatterer placed on a plate behaves as a barrier confining the low-energy eigenfunctions. Although it has been proved that such a localization property has to diminish as the mode number increases, we can increase the number of localized modes by elongating the plate. Note that the lowest eigenfunction should be excluded from this phenomenon since the point scatterer attracts its amplitude when the corresponding eigenvalue is large and negative regardless of the eccentricity of the plate.

\begin{acknowledgements}
The author is greatly indebted to Maciej Zworski for introducing the topic with inspiring discussions. The author was supported by Samsung Scholarship.
\end{acknowledgements}

\nocite{*}
\bibliographystyle{apsrev4-1}
\bibliography{sebabib}

\end{document}